\documentclass[letter]{aa} 
\usepackage{graphicx}
\usepackage{txfonts}
\usepackage{epsfig}
\begin{document}
   \title{AGILE detection of intense gamma-ray emission from the blazar PKS 1510-089}

\author{G.~Pucella\inst{1}, V.~Vittorini\inst{1,2}, F.~D'Ammando\inst{1,3}, M.~Tavani\inst{1,3}, C.~M.~Raiteri\inst{4}, M.~Villata\inst{4}, A.~Argan\inst{1}, G.~Barbiellini\inst{5}, F.~Boffelli\inst{6,7}, A.~Bulgarelli\inst{8}, P.~Caraveo\inst{9}, P.~W.~Cattaneo\inst{6}, A.~W.~Chen\inst{2,9}, V.~Cocco\inst{1}, E.~Costa\inst{1}, E.~Del Monte\inst{1}, G.~De Paris\inst{1}, G.~Di Cocco\inst{8}, I.~Donnarumma\inst{1}, Y.~Evangelista\inst{1}, M.~Feroci\inst{1}, M.~Fiorini\inst{8}, T.~Froysland\inst{2,3}, F.~Fuschino\inst{8}, M.~Galli\inst{10}, F.~Gianotti\inst{8}, A.~Giuliani\inst{9}, C.~Labanti\inst{8}, I.~Lapshov\inst{1}, F.~Lazzarotto\inst{1}, P.~Lipari\inst{11}, F.~Longo\inst{5}, M.~Marisaldi\inst{8}, S.~Mereghetti\inst{9}, A.~Morselli\inst{12}, L.~Pacciani\inst{1}, A.~Pellizzoni\inst{9}, F.~Perotti\inst{9}, P.~Picozza\inst{12}, M.~Prest\inst{13}, M.~Rapisarda\inst{14}, A.~Rappoldi\inst{6}, P.~Soffitta\inst{1}, M.~Trifoglio\inst{8}, A.~Trois\inst{1}, E.~Vallazza\inst{5}, S.~Vercellone\inst{9}, A.~Zambra\inst{1}, D.~Zanello\inst{11}, L.~A.~Antonelli\inst{15}, S.~Colafrancesco\inst{15}, S.~Cutini\inst{15}, D.~Gasparrini\inst{15}, P.~Giommi\inst{15}, C.~Pittori\inst{15}, F.~Verrecchia\inst{15}, L.~Salotti\inst{16}, M.~F.~Aller\inst{17}, H.~D.~Aller\inst{17}, D.~Carosati\inst{18}, V.~M.~Larionov\inst{19}, R.~Ligustri\inst{20}}



\institute{$^{ 1}$ INAF/IASF--Roma, Via del Fosso del Cavaliere 100, I-00133 Roma, Italy \\ 
           $^{ 2}$ CIFS--Torino, Viale Settimio Severo 3, I-10133 Torino, Italy \\ 
           $^{ 3}$ Dip. di Fisica, Univ. di Roma ``Tor Vergata'', Via della Ricerca Scientifica 1, I-00133 Roma, Italy \\ 
           $^{ 4}$ INAF, OATo, Via Osservatorio 20, I-10025 Pino Torinese (Torino), Italy \\ 
           $^{ 5}$ Dip. di Fisica and INFN Trieste, Via Valerio 2, I-34127 Trieste, Italy \\ 
           $^{ 6}$ INFN--Pavia, Via Bassi 6, I-27100 Pavia, Italy \\ 
           $^{ 7}$ Dip. di Fisica Nucleare e Teorica, Univ. di Pavia, Via Bassi 6, I-27100 Pavia, Italy \\ 
           $^{ 8}$ INAF/IASF--Bologna, Via Gobetti 101, I-40129 Bologna, Italy \\ 
           $^{ 9}$ INAF/IASF--Milano, Via E.~Bassini 15, I-20133 Milano, Italy \\ 
           $^{10}$ ENEA--Bologna, Via dei Martiri di Monte Sole 4, I-40129 Bologna, Italy \\ 
           $^{11}$ INFN--Roma ``La Sapienza'', Piazzale A. Moro 2, I-00185 Roma, Italy \\ 
           $^{12}$ INFN--Roma ``Tor Vergata'', Via della Ricerca Scientifica 1, I-00133 Roma, Italy \\ 
           $^{13}$ Dip. di Fisica, Univ. dell'Insubria, Via Valleggio 11, I-22100 Como, Italy \\ 
           $^{14}$ ENEA--Roma, Via E. Fermi 45, I-00044 Frascati (Roma), Italy \\ 
           $^{15}$ ASI--ASDC, Via G. Galilei, I-00044 Frascati (Roma), Italy \\ 
           $^{16}$ ASI, Viale Liegi 26, I-00198 Roma, Italy \\ 
           $^{17}$ Department of Astronomy, University of Michigan, U. S.\\ 
           $^{18}$ Armenzano, Astronomical Observatory, I-06083 Assisi (Perugia), Italy \\ 
           $^{19}$ Astron. Inst., St-Petersburg State University, Russia \\ 
           $^{20}$ Circolo AStrofili Talmassons, Via Cadorna 57, I-33030 Talmassons (Udine), Italy} 
           
\offprints{G. Pucella, \email{gianluca.pucella@iasf-roma.inaf.it} }

   \date{received; accepted}

 
  \abstract
%
   {We report the detection by the AGILE {{\it (Astro-rivelatore Gamma a Immagini
  LEggero)}} satellite of an intense gamma-ray
  flare from the source AGL J1511-0909, associated with the powerful quasar PKS 1510-089, during ten days of observations from 23 August to 1 September 2007.}
%
   {During the observation period, the source was in optical decrease
   following
   a flaring event monitored by the GLAST-AGILE Support Program (GASP) of the
   Whole Earth Blazar Telescope (WEBT). The simultaneous
   gamma-ray, optical, and radio coverage allows us to study the spectral energy distribution and the theoretical models based on the
   synchrotron and inverse Compton (IC) emission mechanisms.}
%
   {AGILE observed the source with its two co-aligned imagers, the Gamma-Ray
   Imaging Detector and the hard X-ray imager Super-AGILE
   sensitive in the 30~MeV $\div$ 50~GeV and 18 $\div$ 60 keV bands, respectively.}
%
   {Between 23 and 27 August 2007, AGILE detected gamma-ray emission from PKS 1510-089 when this source was located $\sim 50^\circ$ off-axis, with an average flux of $(270 \pm 65) \times 10^{-8}$ photons cm$^{-2}$ s$^{-1}$ for photon energy above 100 MeV. In the following period, 28 August - 1 September, after a satellite re-pointing, AGILE detected the source at $\sim 35^\circ$ off-axis, with an average flux (E $>$ 100 MeV) of $(195 \pm 30) \times 10^{-8}$ photons cm$^{-2}$ s$^{-1}$. No emission was detected by Super-AGILE, with a 3-$\sigma$ upper limit of 45 mCrab in 200 ksec.} 
%
   {The spectral energy distribution is modelled with a homogeneous
   one-zone synchrotron self Compton (SSC) emission plus contributions
   by external photons: the SSC emission contributes primarily to the
   X-ray band, whereas the contribution of the IC from the external disc and the broad line region match the hard gamma-ray spectrum observed.} 
\authorrunning{G. Pucella et al.}
\titlerunning{AGILE detection of PKS 1510-089}
\maketitle
%


%
  \section{Introduction}

The radio source PKS 1510-089 was first identified optically as a quasar with an ultraviolet excess, a visual magnitude of 16.5 (Bolton $\&$ Ekers 1966), and a redshift of z = 0.361 measured from its emission-line spectrum (Burbidge $\&$ Kinnan 1966). PKS 1510-089 is a radio-loud highly polarized quasar (HPQ) belonging to the class of the flat spectrum radio quasar (FSRQ) in terms of its spectral energy distribution. Its radiative output is dominated by the gamma-ray component, while its synchrotron emission peaks around IR frequencies below a pronunced UV bump, presumably due to the thermal emission from the accretion disc (Malkan $\&$ Moore 1986, Pian $\&$ Treves 1993). 

PKS 1510-089 has been extensively observed in X-rays by $EXOSAT$ (Singh, Rao
$\&$ Vahia 1990, Sambruna et al. 1994), $GINGA$ (Lawson $\&$ Turner 1997),
$ROSAT$ (Siebert et al. 1998), $ASCA$ (Singh, Shrader $\&$ George 1997) and
$Chandra$ (Gambill et al. 2003). The observed X-ray spectrum was very flat in
the 2 $\div$ 10 keV band (photon index $\Gamma$ $\simeq$ 1.3), but steepened
in the $ROSAT$ bandpass below 2 keV ($\Gamma$ $\simeq$ 1.9). Observations by
$BeppoSAX$ (Tavecchio et al. 2000) confirm the presence of a soft X-ray excess
below 1 keV. A similar soft excess has been detected in other blazar{\bf s}
such as 3C273, 3C279, and 3C454.3, and the origin of the soft X-ray excess is still an open issue. 

PKS 1510 was recently observed by $Suzaku$ in August 2006 over approximately
three days, and the campaign continued with {\it Swift} monitoring over
18 days (Kataoka et al. 2007). {\it Swift}-XRT observations reveal significant
spectral evolution of the X-ray emission on timescales of one week: the X-ray spectrum becomes harder as the source gets brighter. 

Gamma-ray emission from PKS 1510-089 was detected by the EGRET instrument on board $CGRO$ with a integrated flux above 100 MeV between (13 $\pm$ 5) and (49 $\pm$ 18) $\times$ 10$^{-8}$ photons cm$^{-2}$ s$^{-1}$ and an energy spectrum modelled with a power law with a photon index $\Gamma =$ 2.47 $\pm$ 0.21. In this Letter we present the analysis of the AGILE data obtained during the PKS 1510-089 observations from 23 August 2007 to 1 September 2007.

%


%
  \section{AGILE observation of PKS 1510-089} \label{1510:pointing}

\begin{figure}
 \centering
  \includegraphics[width=5.8cm, height=5.8cm]{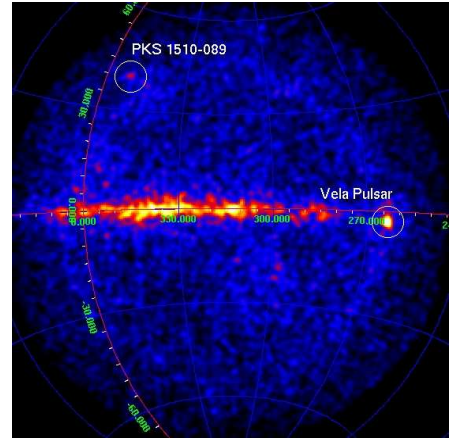}
   \caption{Gaussian-smoothed count map ($\sim 140^\circ \times 140^\circ$) in Galactic coordinates integrated over 
            the observing period 2007 August 23 - August 27. The circles are located at the 
            PKS 1510-089 and Vela Pulsar coordinates. Only photons
            with energy greater than 100 MeV have been included.}
   \label{1510GRID1}
\end{figure}

The AGILE scientific Instrument (Tavani et al.\ 2008) is very compact and
combines four active detectors yielding broad-band coverage from hard X-rays
to gamma rays. 

The Gamma-Ray Imaging Detector (GRID) consists of a combination of a silicon
tracker (Prest et al. 2003, Barbiellini et al. 2001), a non-imaging Cesium
Iodide Mini-Calorimeter (MCAL; Labanti et al. 2006) positioned under the
silicon tracker and sensitive in the 0.3 $\div$ 100~MeV energy band, and a
segmented anticoincidence system (ACS) made of a plastic shield which surrounds all
active detectors (Perotti et al. 2006). A co-aligned coded-mask hard X-ray
imager  (SuperAGILE; Costa et al. 2001, Feroci et al. 2007) ensures coverage
in the 18 $\div$ 60~keV energy band. 

The GRID has a field of view of $\sim$ 2.5 sr, an angular resolution of 1.2$^{\circ}$ at 400 MeV (68$\%$ cont. radius), an effective area of $\sim$ 500
  cm$^{2}$ above 100 MeV, and an energy resolution $\Delta$E$/$E $\sim$ 1 at
  400 MeV. The silicon tracker and the on-board trigger logic are optimized for gamma-ray imaging in the 30~MeV $\div$ 50~GeV energy band (Argan et al. 2004). 

The AGILE observations of the PKS 1510-089 were performed from 23 August 2007
  12:00 UT to 1 September 2007 12:00 UT, for a total of 84 hours of effective
  exposure time. In the first period, between 23 and 27 August, the source was located $\sim 50^\circ$ off the AGILE pointing direction. In the second period, between 28 August and 1 September, after a satellite re-pointing, the source was located at $\sim 35^{\circ}$ off-axis. 

\begin{figure}
 \centering
  \includegraphics[width=5.8cm, height=5.8cm]{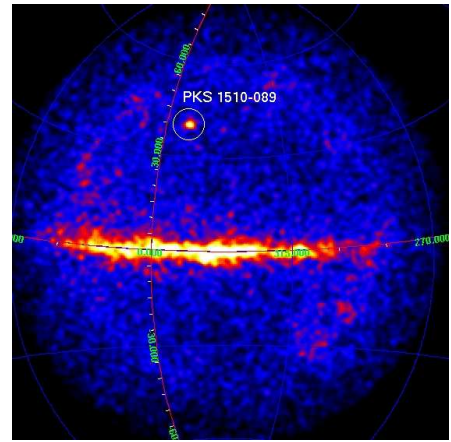}
   \caption{Gaussian-smoothed count map ($\sim 140^\circ \times 140^\circ$) in Galactic coordinates integrated over 
            the observing period 2007 August 28 - September 1. The circle is located at the 
            PKS 1510-089 coordinates. Only photons with energy greater
            than 100 MeV have been included.}
   \label{1510GRID2}
\end{figure}
%


%
\section{Data reduction and analysis} \label{1510:dataanal}

Level--1 AGILE-GRID data were analysed using the AGILE Standard Analysis
Pipeline. The first step is to align all data times to terrestrial time (TT),
and it performs preliminary calculations. In the second step, an ad-hoc
implementation of the Kalman Filter technique is used for track identification
and event-direction reconstruction in detector coordinates. Subsequently, a
quality flag is assigned to each GRID event: (G), (P), (S), and (L), depending
on whether it is recognised as a confirmed gamma-ray event, a charged
particle event, a single-track event, or its nature is uncertain,
respectively. The (L) event class includes events typically affected by an
order of magnitude higher particle contamination than (G). The single-track
(S) event class includes events for which only one track is reconstructed in the two orthogonal views of the tracker. Then, an AGILE log-file is created,
containing all the information relevant to computating the exposure and
live time. The third step is to create the AGILE event files, excluding events flagged as particles. This step also reconstructs the event direction in sky coordinates. 

Once the above steps are completed, the AGILE Scientific Analysis Package can be run. Counts, exposure, and Galactic background gamma-ray maps are created with a bin-size of $0.5^{\circ} \times 0.5^{\circ}$\, for photons with energy over 100 MeV. To reduce the particle background contamination, we selected only events flagged as confirmed gamma-ray events, and all events collected during the South Atlantic Anomaly were rejected. 

We also rejected all the gamma-ray events whose reconstructed directions form angles with the satellite-Earth vector smaller than 80$^{\circ}$, reducing the gamma-ray Earth albedo contamination by excluding regions within $\sim 10^{\circ}$ from the Earth limb. We ran the AGILE maximum likelihood procedure (ALIKE) on the whole observing period, in order to obtain the average flux in the gamma-ray band.

\begin{figure}
 \vspace{-0.4cm}
  \centering
   \includegraphics[width=9.0cm]{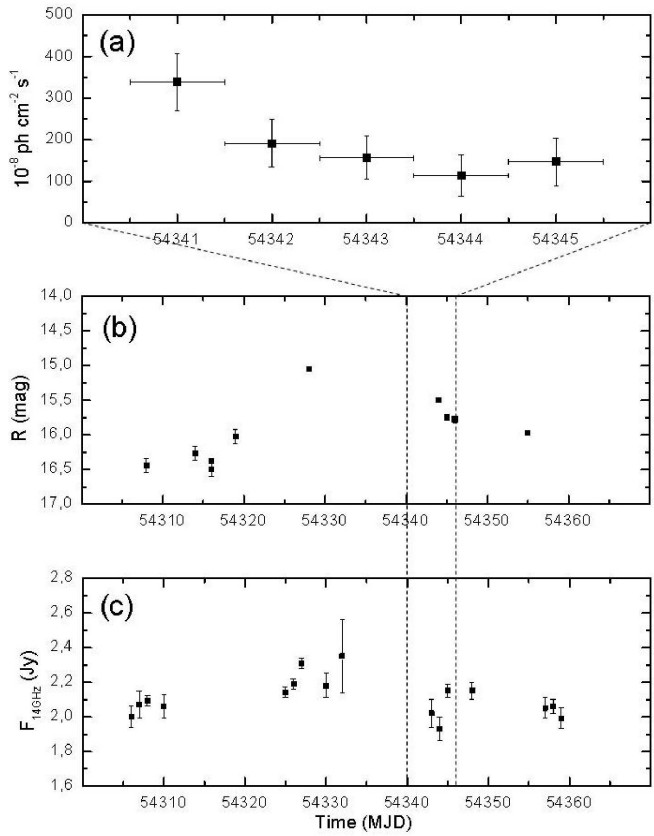}
    \vspace{-0.4cm}
     \caption{a) AGILE-GRID gamma-ray light curve, with a 1-day resolution,
              for the observation period 2007 August 28 - September 1, 
              for photons with E $>$ 100 MeV in units of 10$^{-8}$ photons cm$^{-2}$ s$^{-1}$. 
              b) R-band optical light curve as observed by the GASP of the
              WEBT for the observation period 2007 July 26 - September 11. 
              c) GASP radio light curve (from UMRAO) at 14.5 GHz for the
              observation period 2007 July 24 - September 15. }
  \label{1510lc}
\end{figure}

\begin{figure}
 \vspace{-0.4cm}
  \centering
   \includegraphics[width=7.5cm]{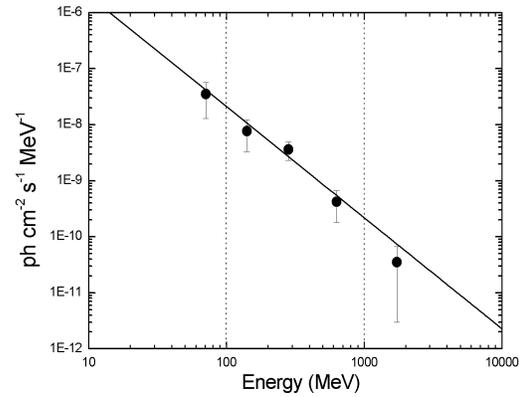}
   \vspace{-0.4cm}
    \caption{AGILE average gamma-ray spectrum of PKS 1510-089 for the observation period 28 August - 1 September 2007.} 
  \label{1510spectrum}
\end{figure}

%


%
 \section{Results} \label{1510:results}

Figure \ref{1510GRID1} shows a Gaussian-smoothed count map in Galactic
coordinates integrated over the observing period 2007 August 23 - 27
for photon energies over 100 MeV. In this period, AGILE detected
gamma-ray emission from a position consistent with the quasar PKS 1510-089 at
a significance level of 5.6-$\sigma$ as derived from a maximum likelihood
analysis using the PKS 1510-089 radio position $(l=351.29^\circ,
b=40.14^\circ)$. During this period the source was located $\sim 50^\circ$ off
the AGILE pointing direction. Thanks to the large field of view of AGILE,
during this period, the satellite simultaneously also detected the gamma
activity of the source Vela Pulsar $(l=263.55^\circ, b=-2.79^\circ)$, about 90
degrees from PKS 1510-089. In addition, the two sources were characterised by
an almost equal angular distance from the AGILE pointing direction. This
allowed us to obtain an estimate of average flux of PKS 1510-089 through a
direct calibration with the flux of the Vela Pulsar. In this way, the average
flux (E $>$ 100~MeV) estimated for this first period was $(270 \pm 65) \times
10^{-8}$ photons cm$^{-2}$ s$^{-1}$. The reduced effective area and the
consequently reduced count statistics for these large off-axis angles makes it
difficult to create a light curve and an average energy spectrum. 

\begin{figure*}
\vspace{-0.4cm}
 \centering 
  \includegraphics[width=14cm, height=7.5cm] {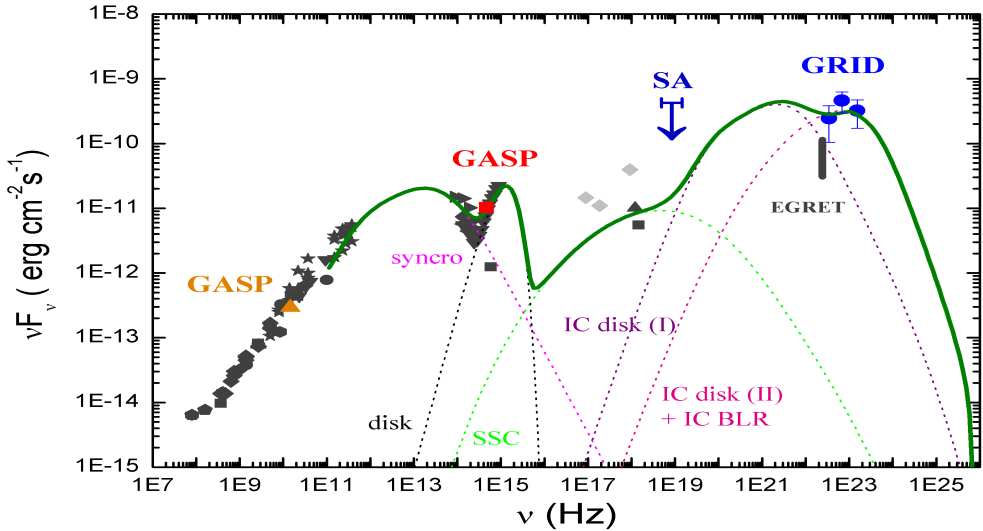}
\vspace{-0.4cm}
   \caption{Spectral Energy Distribution of PKS 1510-089 for the GRID
            observation period 28 August - 1 September 2007 (blue dots), 
            including simultaneous GASP optical (red square) and radio
            (orange triangle) data and the 3-$\sigma$ Super-AGILE upper limit. 
            Non-simultaneous historical data (from 1969 to 2007) taken
            from NASA Extragalactic Database (NED) and Kataoka et
            al. (2007) over the entire electromagnetic spectrum are
            represented in dark grey and light grey, respectively. }
  \label{1510_SED}
\end{figure*}

Figure \ref{1510GRID2} shows a Gaussian-smoothed count map in Galactic
coordinates integrated over the observing period 2007 August 27 --
September 1 for photon energies higher than 100 MeV. In this period the
satellite detected the source at a significance level of 10.6-$\sigma$. With
the likelihood method, the best position for the gamma-ray emission above 100
MeV is $(l=350.89^\circ, b=40.34^\circ)$, with an average
flux (E $>$ 100~MeV) over the period of $(195 \pm 30) \times 10^{-8}$ photons
cm$^{-2}$s$^{-1}$. The AGILE 95$\%$ maximum-likelihood contour level has a
semi-major axis $a$ = 0.43$^\circ$ and semi-minor axis $b$ = 0.09$^\circ$. The
overall AGILE error circle, taking both statistical and systematic effects
into account, has a radius $r$ = 0.53$^\circ$. During this period the source was observed at $35^{\circ}$
off-axis. In Fig. \ref{1510lc} the gamma-ray light curve for
this period with a 1-day resolution, the ${\it R}$-band optical light curve as
observed by the GASP of the WEBT for the observation period between 26 July
and 11 September 2007, and the GASP radio light curve (from UMRAO) at 14.5 GHz
for the observation period between 24 July and 15 September 2007 \footnote{The
  optical and radio data presented in this paper are stored in the GASP-WEBT
  archive (http://www.oato.inaf.it/blazars/webt). For questions regarding
  their availability, please contact the WEBT President Massimo Villata.}
  are reported.
 Figure \ref{1510spectrum} reports the average energy spectrum for this second period,
when only three energy bins were taken into account for the fit: 100 $\div$ 200
MeV, 200 $\div$ 400 MeV, 400 $\div$ 1000 MeV. The energy bins 50 $\div$ 100
MeV and 1000 $\div$ 3000 MeV have not be used in the spectral analysis
waiting for on-flight calibration finalization that will include these
two energy channels.

A simple power-law model can fit the data. We obtain a photon index $\Gamma =$
1.98 $\pm$ 0.27 with the weighted least squares method. 

Super-AGILE observed PKS 1510-089 for a total on-source effective
exposure time of 200 ks. The source was not detected above 5-$\sigma$ by the
Super-AGILE Iterative Removal Of Sources (IROS) applied to the  image, in
the 20--60~keV energy range. A 3-$\sigma$ upper limit of 45 mCrab was obtained
from the observed count rate by a study of the background fluctuations at the
position of the source and a simulation of the source and background
contributions with IROS.

Finally, in Fig. 
\ref{1510_SED} the spectral energy distribution is shown for the GRID
observation period 2007 August 28 - September 1, including simultaneous GASP optical and radio data and the 3-$\sigma$ Super-AGILE upper limit. Also non-simultaneous historical data over the entire electromagnetic spectrum are represented. 

%

%

\section{Discussion} \label{1510:discussion}

During the AGILE observation period, the PKS 1510-089 optical flux appears to
be decreasing in the range ${\it R}\sim$ 15.5 $\div$ 15.8, following a bright
state that reached at least ${\it R}$ = 15.0 (see Fig. \ref{1510lc}). The
contemporaneous gamma-ray flux decrease of about a factor 3 suggests that the
two flux variations may be correlated. In agreement with Kataoka et al. (2007),
  in order to model the spectral energy
distribution we used a homogeneous one-zone synchrotron self-Compton (SSC)
model, plus the contribution of external seed photons originating both
  from the accretion disc and the broad line region (BLR). We obtained a
  representative fit for the spectral energy distribution with input parameters
  similar to those chosen by Kataoka et al. (2007). We consider a
  relativistic moving spherical blob of radius R = 2.5 $\times$ 10$^{16}$ cm and an electron energy distribution described by
a double power law:
\begin{equation}
n_{e}(\gamma)=\frac{K\gamma_{b}^{-1}}{(\gamma /\gamma_{b})^{p_1}+(\gamma
/\gamma_{b})^{p_2}}
\end{equation}
for electron Lorentz factor 40 $<\gamma<$ 4 $\times$$10^{3}$ with spectral indices pre- and
post-break $p_1=2.0$ and $p_2=4.5$, a normalization factor K = 80 cm$^{-3}$ and
the break energy Lorentz factor $\gamma_{b}$ = 400. We assumed a
magnetic field B = 3 Gauss and a Doppler factor $\delta=9$ for the blob.


In order to interpret our gamma-ray data, an accretion disc characterised by
  a blackbody spectrum with a luminosity of
$10^{46}$erg s$^{-1}$ at 0.1 pc from the blob is assumed as the source of external
  target photons. The inverse Compton
(IC) contribution from the disc is calculated up to the second order, but it
is not enough to account for the high gamma-ray state observed by AGILE. The
addition of the IC emission from a BLR, represented by a spherical
  layer extending between 0.1 pc and 0.4 pc from the central black hole,
  reprocessing a 10$\%$ of the irradiating continuum can explain the high
  state observed by AGILE compared to the historical EGRET observations (see
  Fig. \ref{1510_SED}), and it reflects on the different photon index
  obtained in the AGILE and EGRET observations. In this model, the SSC
  emission  primarily contributes to the X-ray band, whereas the IC
  contribution from the BLR can explain the observed hard gamma-ray spectrum.

%

\begin{acknowledgements}

The AGILE Mission is funded by the Italian Space Agency (ASI) with scientific
and programmatic participation by the Italian Institute of Astrophysics (INAF)
and the Italian Institute of Nuclear Physics (INFN). This research has made
use of the NASA/IPAC Extragalactic Database (NED) which is operated by the Jet
Propulsion Laboratory, California Institute of Technology, under contract with the National
Aereonautics and Space Administration.

\end{acknowledgements}
%

%

\end{document}